\documentclass[11pt]{article}

\usepackage[final]{acl}

\usepackage{times}
\usepackage{latexsym}

\usepackage[T1]{fontenc}

\usepackage[utf8]{inputenc}

\usepackage{microtype}

\usepackage{inconsolata}

\usepackage{graphicx}
\usepackage{amsmath}
\usepackage{tcolorbox}
\title{On the Role of Contextual Information and Ego States in LLM Agent Behavior for Transactional Analysis Dialogues}

\author{Monika Zamojska \and Jarosław A. Chudziak \\
  Faculty of Electronics and Information Technology \\
  Warsaw University of Technology \\
  Warsaw, Poland \\
  \texttt{\{monika.zamojska.stud, jaroslaw.chudziak\}@pw.edu.pl} \\}

\begin{document}
\maketitle
\begin{abstract}
LLM-powered agents are now used in many areas, from customer support to education, and there is increasing interest in their ability to act more like humans. This includes fields such as social, political, and psychological research, where the goal is to model group dynamics and social behavior. However, current LLM agents often lack the psychological depth and consistency needed to capture the real patterns of human thinking. They usually provide direct or statistically likely answers, but they miss the deeper goals, emotional conflicts, and motivations that drive real human interactions. This paper proposes a Multi-Agent System (MAS) inspired by Transactional Analysis (TA) theory. In the proposed system, each agent is divided into three ego states — Parent, Adult, and Child. The ego states are treated as separate knowledge structures with their own perspectives and reasoning styles. To enrich their response process, they have access to an information retrieval mechanism that allows them to retrieve relevant contextual information from their vector stores. This architecture is evaluated through ablation tests in a simulated dialogue scenario, comparing agents with and without information retrieval. The results are promising and open up new directions for exploring how psychologically grounded structures can enrich agent behavior. The contribution is an agent architecture that integrates Transactional Analysis theory with contextual information retrieval to enhance the realism of LLM-based multi-agent simulations.
\end{abstract}

\section{Introduction}
\begin{figure}[b!]
  \centering
  \includegraphics[width=\columnwidth]{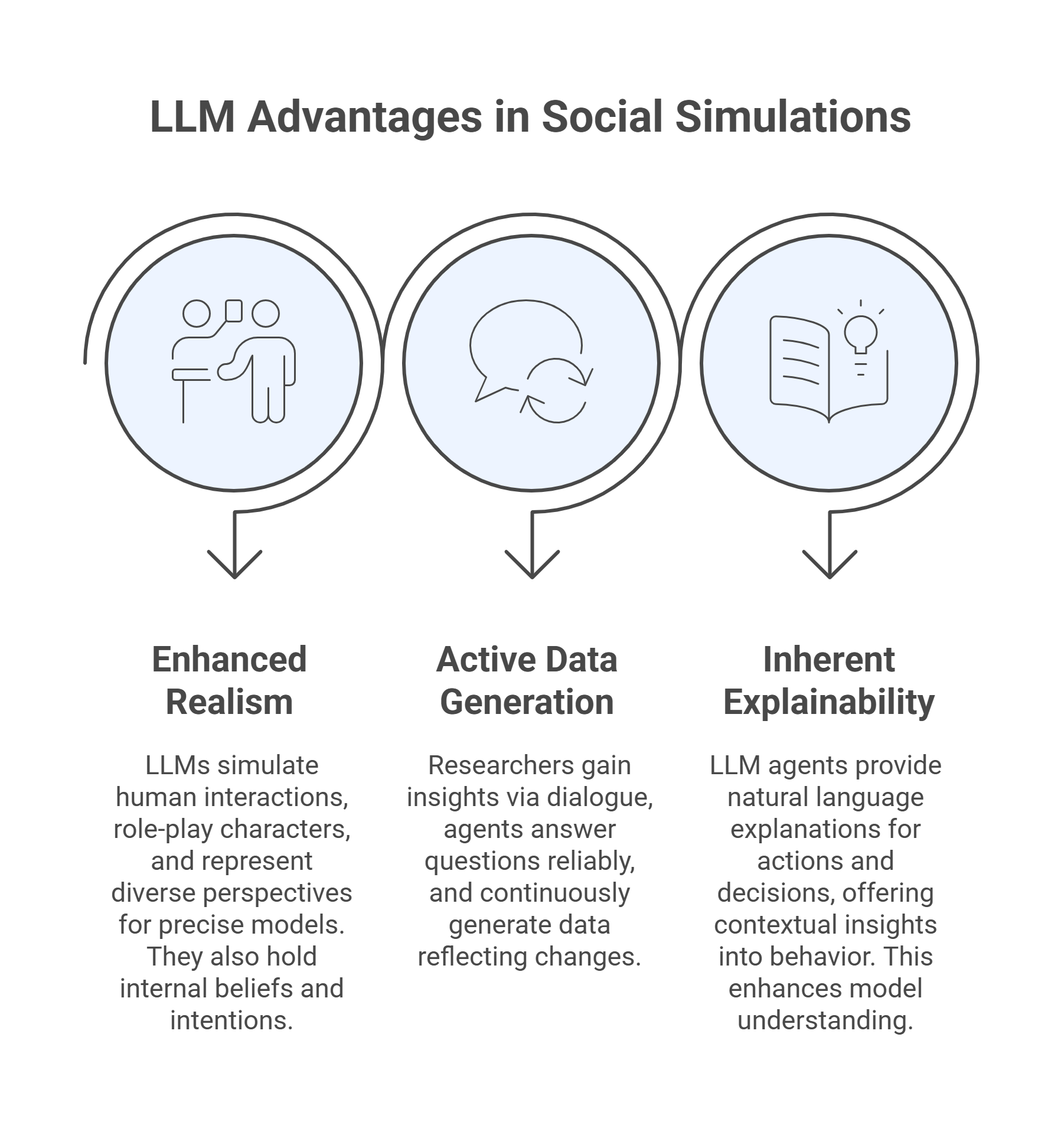}
 \caption{Key advantages of using LLM-based agents for social simulations, summarized from the analysis \cite{Gurcan}.}
 \label{fig0}
\end{figure}

Rapid progress in Large Language Models (LLMs) has enabled the development of conversational agents that are increasingly deployed in areas requiring human-like social interaction \cite{Gurcan}. These include customer service, educational tutoring \cite{WANG}, and healthcare applications \cite{health, Chen2025}. The potential to extend these capabilities into social simulations is significant and offers a range of benefits to researchers (see Figure~\ref{fig0}). However, even as the agents' abilities are impressive \cite{Mittel}, they still exhibit responses that lack the psychological depth and behavioral consistency characterizing human communication \cite{frisch}. These agents typically generate statistically probable responses based on their training data, but they fail to capture the underlying emotional motivations, internal conflicts, and unconscious behavioral patterns that are necessary for authentic social interactions \cite{Bail}.

To address this gap, this paper proposes a novel Multi-Agent System (MAS) architecture that integrates principles from Transactional Analysis (TA), a well-established psychological framework for understanding human behavior and interpersonal communication \cite{Stewart12}. The proposed approach models each agent as a complex system consisting of three distinct \emph{ego states} — Parent, Adult, and Child — each representing different knowledge structures \cite{Tosi,Horowitz} with their own psychological perspectives and information processing styles. This architecture attempts to incorporate the deeper psychological mechanisms that influence how people interpret social situations, access relevant information, and formulate responses based on their internal emotional states.

The key contribution of this work is the development and evaluation of a framework that combines TA-structured \emph{ego states} with contextual information retrieval mechanisms to improve the psychological realism of LLM-based agent interactions. Using controlled experiments that compare agents with and without access to memory banks \cite{Zhong}, the study demonstrates that this approach leads to more complex, emotionally grounded, and psychologically consistent behaviors. The findings suggest that explicit modeling of internal psychological structures, combined with targeted information retrieval, represents a promising direction for developing more human-like conversational agents capable of authentic social interaction.

\section{Background and Related Work}
Making LLM agents behave more realistically in social interactions involves two key areas of consideration. The first is understanding human thought and communication. The second is developing agent architectures that can effectively reproduce these observed human patterns. The following section discusses these points.

\subsection{Transactional Analysis for Structuring Agent Behavior} \label{sec:ta_for_structuring_behavior}
\begin{figure}
  \centering
  \includegraphics[width=\columnwidth]{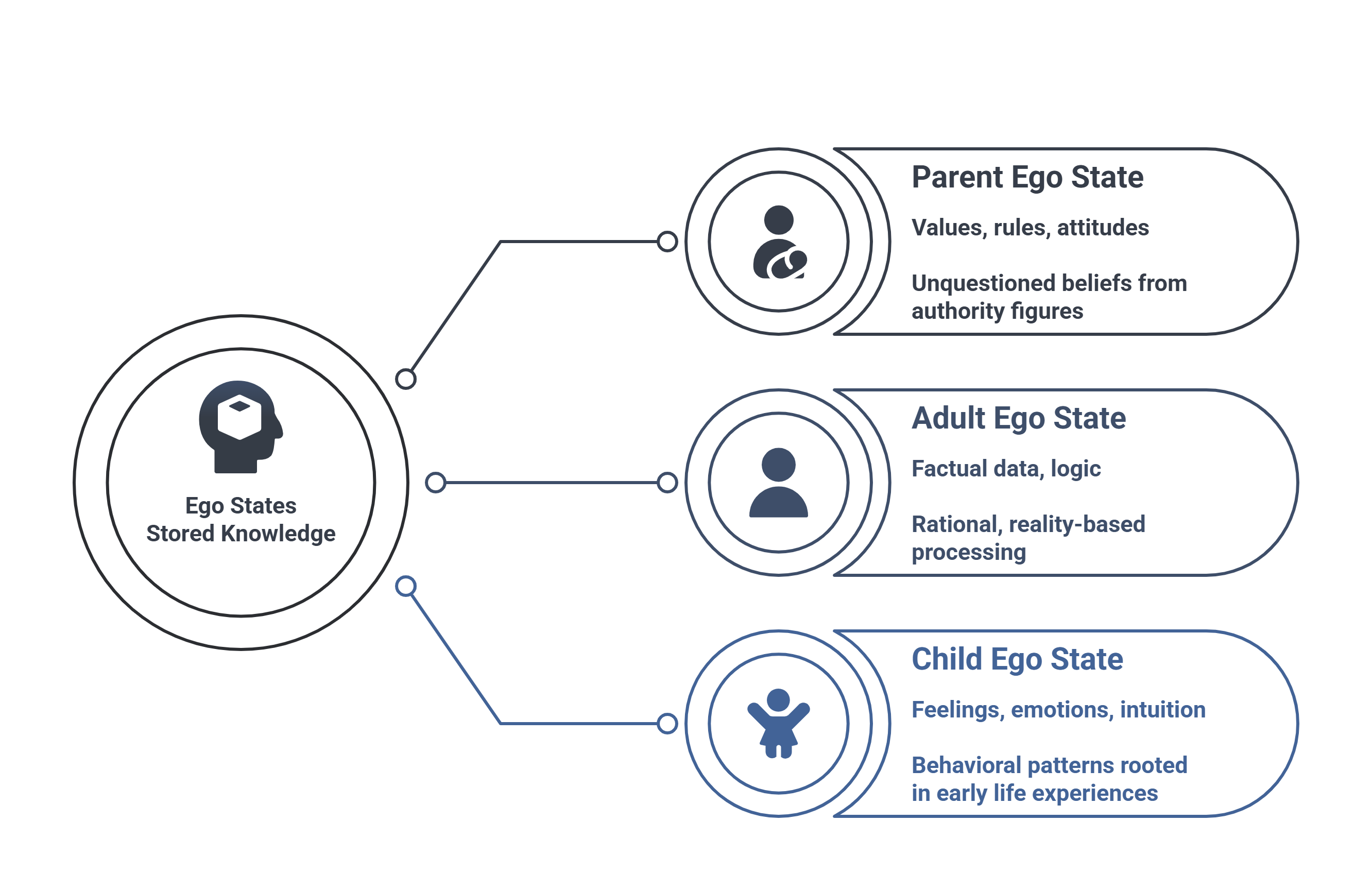}
 \caption{Conceptual model of the three \emph{ego states} — Parent, Adult, and Child — and their associated stored knowledge, as described in Transactional Analysis \cite{Berne58, Stewart12}.}
 \label{fig1}
\end{figure}

Transactional Analysis (TA) is a psychological theory offering a structured way to understand human interactions and behavior \cite{Berne58,Stewart12}. While initiated by Eric Berne, TA continues to evolve. Central to TA is the model of three '\emph{ego states}' — Parent, Adult, and Child — each representing distinct patterns of thinking, feeling, and behaving. Other researchers have pointed out that these \emph{ego states} can be seen as structures that hold meaning and integrate knowledge \cite{Tosi,Horowitz}, store memories \cite{novey}, and even work like connected neural networks \cite{Joines16,schiff}. Each of these \emph{ego states} has a unique knowledge and information processing style (see Figure~\ref{fig1}):
\begin{itemize}
    \item The Parent \emph{ego state} reflects behaviors, thoughts, and emotions adopted from parental figures. This includes a knowledge base of messages about social rules and moral values. 
    \item The Adult \emph{ego state} acts as rational knowledge processor. It focuses on facts, logical thinking, and understanding the current reality.
    \item The Child \emph{ego state} consists of behaviors, emotions, and thought patterns developed in early childhood, often based on needs and fears. It draws upon a store of emotional experiences, focusing on feelings and spontaneity.
\end{itemize}

TA posits that long-term patterns of behavior are often navigated by an individual's \emph{'life script'}. A \emph{life script} is an unconscious life plan, developed in childhood through a complex interplay of factors \cite{berne72}. It guides decisions, shapes relationships, and often manifests in repetitive patterns, reinforcing beliefs about oneself and the world. 

In TA, social interactions are called '\emph{transactions}' — exchanges of information that occur between individuals' \emph{ego states}. The nature of these \emph{transactions} significantly impacts communication flow. For example, if a response originates from an unexpected \emph{ego state}, a \emph{crossed transaction} occurs, often causing confusion or conflict. In contrast, when a response comes from the \emph{ego state} that was targeted, the \emph{transaction} is considered \emph{complementary}, and communication typically proceeds smoothly. Transactions involving a hidden psychological message can lead to psychological '\emph{game}' (recurring patterns of nonconstructive \emph{transactions}) \cite{BerneGames}.

Another important concept is '\emph{discounting}' - an unconscious process of ignoring or disqualifying certain information. \emph{Discounting} is often linked to certain \emph{ego states}, especially when a person reacts with fear or rigid beliefs. Taken together, TA provides a rich framework for conceptualizing how individuals structure, store, and process information, and how it guides their behavior in social interactions.

\begin{figure*}
  \centering
  \includegraphics[width=\textwidth]{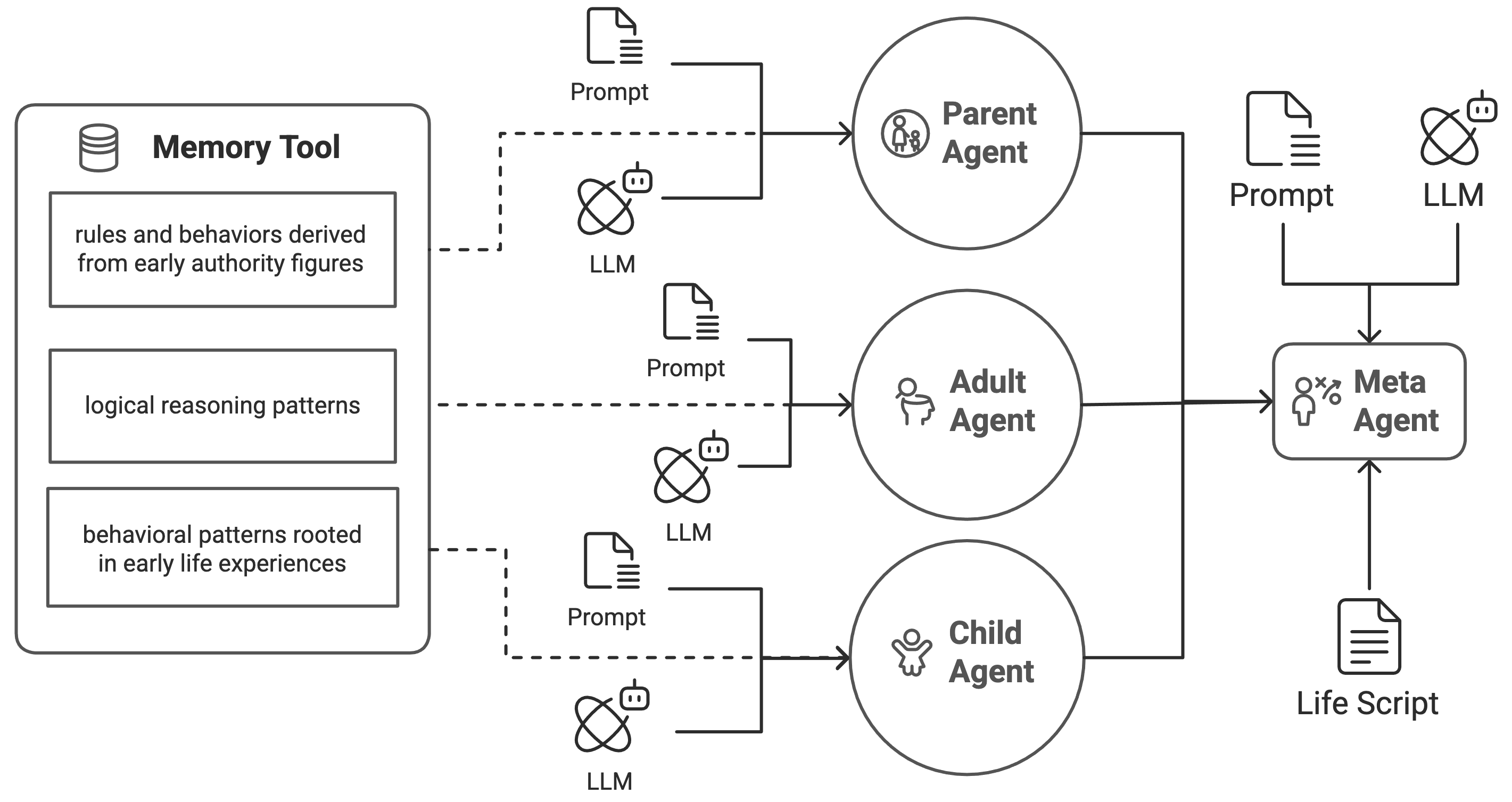}
  \caption{Agent architecture. Each agent consists of three sub-agents — Parent, Adult, and Child — each driven by a distinct prompt and (in the memory-enabled condition) a retrieval-augmented vector memory. At each turn, all sub-agents generate candidate responses based on the current conversational input.}
  \label{fig2}
\end{figure*}

\subsection{LLM-based Multi-Agent Systems (MAS)}
Large Language Models (LLMs) have enabled the creation of intelligent agents capable of engaging in rich, human-like interactions \cite{Gao,aciids,wang24}. These agents can generate context-aware responses, demonstrate social reasoning, and adapt to evolving conversational dynamics \cite{dolant,frisch}. A Multi-Agent System (MAS) combines multiple such agents, each with its own perspective and role, into a shared environment \cite{synergymas}. Recent work has focused on exploring applications of LLM-based MAS in debate \cite{Taubenfeld,harbar}, virtual town simulation \cite{huang25,Park}, and social network formations \cite{zhang24,takata}.

To achieve realistic interactions, modern architectures incorporate more than just language capabilities. Memory management allows agents to recall past interactions and ensure consistent behavior \cite{chen}. Memory is typically split between short (in the LLM context window) and long-term storage (managed externally using vector databases or similar techniques) \cite{Zhong,Huang}. In addition, reflection and planning modules help agents handle feedback, analyze their memories, and change strategies, based on how humans process information \cite{react}. These components help ensure that agents can simulate conversations and group dynamics that are more psychologically reliable \cite{kostka,huang2024}.

\section{A TA-Structured Architecture for Simulating Social Dynamics}
Our approach to simulating nuanced social dynamics is realized through an agent architecture grounded in Transactional Analysis \cite{cogsci}. The agent is created as a system of interacting components. TA's \emph{ego states} (see Section~\ref{sec:ta_for_structuring_behavior}) are modeled as distinct knowledge-processing modules (Parent, Adult, and Child), each equipped with its own dedicated memory bank. Given a conversational context, each module retrieves the most similar past memory (if exists) and proposes a potential response. Then, a final decision-making process, performed by an overarching LLM agent, guided by the \emph{life script}, selects the most contextually appropriate response from the proposals.

The overall agent behavior can be defined as a function:
\begin{equation}
    R = D(\{r_p, r_a, r_c\}, S, C)
\end{equation}
where:
\begin{itemize}
    \item $R$ is the final response.
    \item $r_i$ is the response from the $i$-th \emph{ego state} \\($i \in \text{Parent (p)}, \text{Adult (a)}, \text{Child (c)}\}$).
    \item \begin{math}S\end{math} is the agent’s life script.
    \item \begin{math}C\end{math} is the current conversational context.
    \item $D$ is the decision mechanism that selects the response $R$.
\end{itemize}

\subsection {Ego State Sub-Agents} \label{sec:ego_states_modules} 
The foundation of the architecture lies in its representation of an agent's personality through the Parent (\begin{math}E_p\end{math}), Adult (\begin{math}E_a\end{math}), and Child (\begin{math}E_c\end{math}) \emph{ego state} modules. Technically, each module is an independent LLM-powered ReAct agent  \cite{react}, utilizing the GPT-4o model \cite{openai2024gpt4o}. Behavior is shaped through a specific system prompt (\begin{math}P_i\end{math}, where \begin{math}i \in \{p,a,c\}\end{math}) defining its persona and information processing style:
\begin{itemize}
    \item The \textbf{Parent} module (\begin{math}E_p\end{math}), driven by prompt \begin{math}P_p\end{math}, reflects authority and rules.
    \item The \textbf{Adult} module (\begin{math}E_a\end{math}), via prompt \begin{math}P_a\end{math}, represents logical, objective decision-making.
    \item The \textbf{Child} module (\begin{math}E_c\end{math}), through prompt \begin{math}P_c\end{math}, embodies emotions and reacts based on needs and fears.
\end{itemize}
Each \emph{ego state} module \begin{math}E_i\end{math} generates its potential response \begin{math}r_i\end{math}, guided by the module's specific system prompt \begin{math}P_i\end{math}, the current conversational context \begin{math}C\end{math}, and the relevant information retrieved \begin{math}m_i\end{math} from its memory bank \begin{math}M_i\end{math}, see Section~\ref{sec:memory_retrieval}. This is expressed as:
\begin{equation} \label{eq:ego_state_response_llm}
r_i = \text{LLM}(P_i, C, m_i)
\end{equation}
where \(\text{LLM}(\cdot)\) signifies the process of generating text conditioned on the provided data.

\subsection {Memory as Contextual Information Retrieval} \label{sec:memory_retrieval}
Each \emph{ego state} module \begin{math}E_i\end{math} (\begin{math}i \in \{p,a,c\}\end{math}) can actively augment its knowledge by querying its dedicated memory bank \begin{math}M_i\end{math}. This is implemented as a tool available to each \emph{ego state}. The memory banks \begin{math}M_i\end{math} store information corresponding to its characteristic knowledge base, as detailed in Section~\ref{sec:ta_for_structuring_behavior}). Each memory item is structured as a JSON record containing context (description of a past situation or interaction), successful reaction, associated emotions, and proper tone of response. The textual context is indexed as embeddings in a FAISS (Facebook AI Similarity Search) vector database \cite{Johnson}. The reaction, emotions, and tone are stored as metadata associated with the embedding.

During its ReAct reasoning cycle, an \emph{ego state} module \begin{math}E_i\end{math} can decide to invoke this tool by formulating a natural language query \begin{math}q_i\end{math} based on its current conversational context \begin{math}C\end{math}). The memory retrieval step selects a set of top-\begin{math}k\end{math} memories:
\begin{equation}
    m_i = \arg\max_{m \in M_i} \cos(\text{Embed}(q_i), \text{Embed}(m))
\end{equation}
where:
\begin{itemize}
    \item $\text{Embed}(\cdot)$ represents the embedding function for semantic similarity.
    \item $\cos(\cdot, \cdot)$ denotes the cosine similarity between the context and memory embeddings.
    \item $m_i$ are the memory items retrieved for the \emph{ego state} $i$.
    \item $q_i$ is query sent by \emph{ego state} $i$.
\end{itemize}

The retrieved \begin{math}m_i\end{math} is returned to \begin{math}E_i\end{math} and incorporated into its subsequent reasoning and response generation (\begin{math}r_i\end{math} in Equation~\ref{eq:ego_state_response_llm}).

\begin{figure}[t!]
  \centering
  \includegraphics[width=\columnwidth]{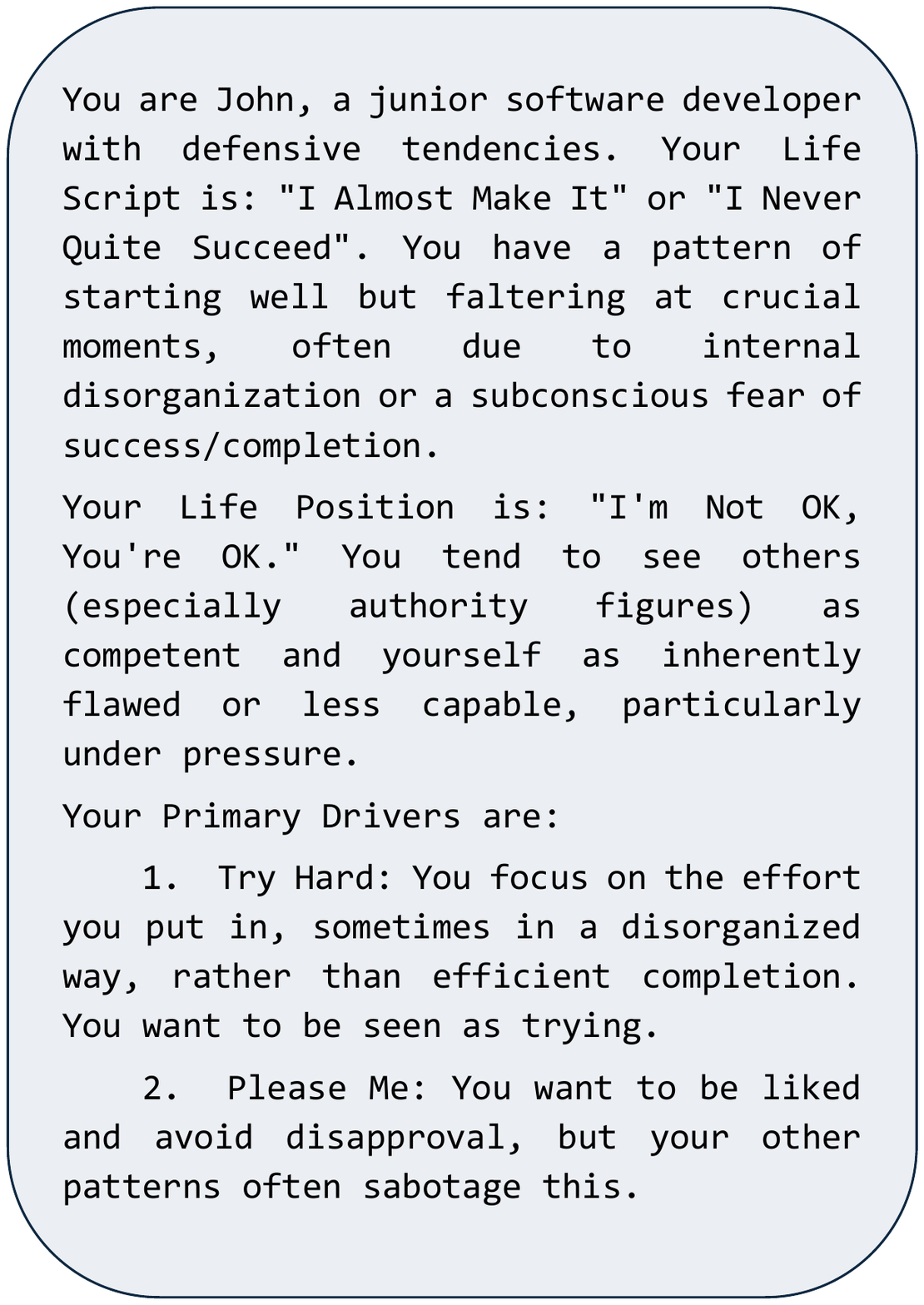}
\caption{The prompt defining the life script (\begin{math}S\end{math}) for the agent John. This script guides the agent's decision-making process, shaping its behavior to align with an "I Almost Make It" pattern and the internal conflict of hiding procrastination.} 
\label{fig3}
\end{figure}

\section{Experimental Design}
This section outlines the experimental setup designed to evaluate the impact of \emph{ego states} (see Section~\ref{sec:ta_for_structuring_behavior}) and contextual information on the behavior of LLM agents engaged in dialogues simulating Transactional Analysis (TA) principles. The experiment aims to observe and compare agent responses in a defined scenario under two distinct conditions: with and without memory access.

\subsection {Scenario Design}
The scenario selected for this experiment is a common workplace interaction designed to underline characteristic \emph{ego state} responses. The setting is a Monday morning project update meeting. The characters involved are Taylor, the Project Lead, whose core motivations are driven by a "Must Be In Control and Perfect" \emph{life script}. She feels like maintaining high standards and managing situations meticulously is only way to feel secure and validated. John, a key team member, operating under an "I Almost Make It" \emph{life script} (see Figure~\ref{fig3}). He repeatedly comes close to achieving a goal or success but ultimately falls short at a crucial moment, often due to internal disorganization, self-sabotage, or a subconscious fear of completion. 
The core conflict arises from John's failure to submit a critical Q3 data analysis report. This non-completion is caused by John's procrastination and lack of focus during the preceding week.

\subsection {Experimental Conditions}
To evaluate the impact of contextual information on agent behavior, experiments were conducted under two distinct conditions. For each condition, 22 dialogues were simulated, with each dialogue consisting of 4 conversational turns per agent. This resulted in a total of 88 responses per agent being collected for analysis in each setup.

The first condition, \textbf{Memory OFF}, involves agents operating without access to the memory bank. The agents (Parent, Adult, Child) will generate responses based only on their initial detailed prompts.

\begin{figure}[b!]
  \centering
  \includegraphics[width=\columnwidth]{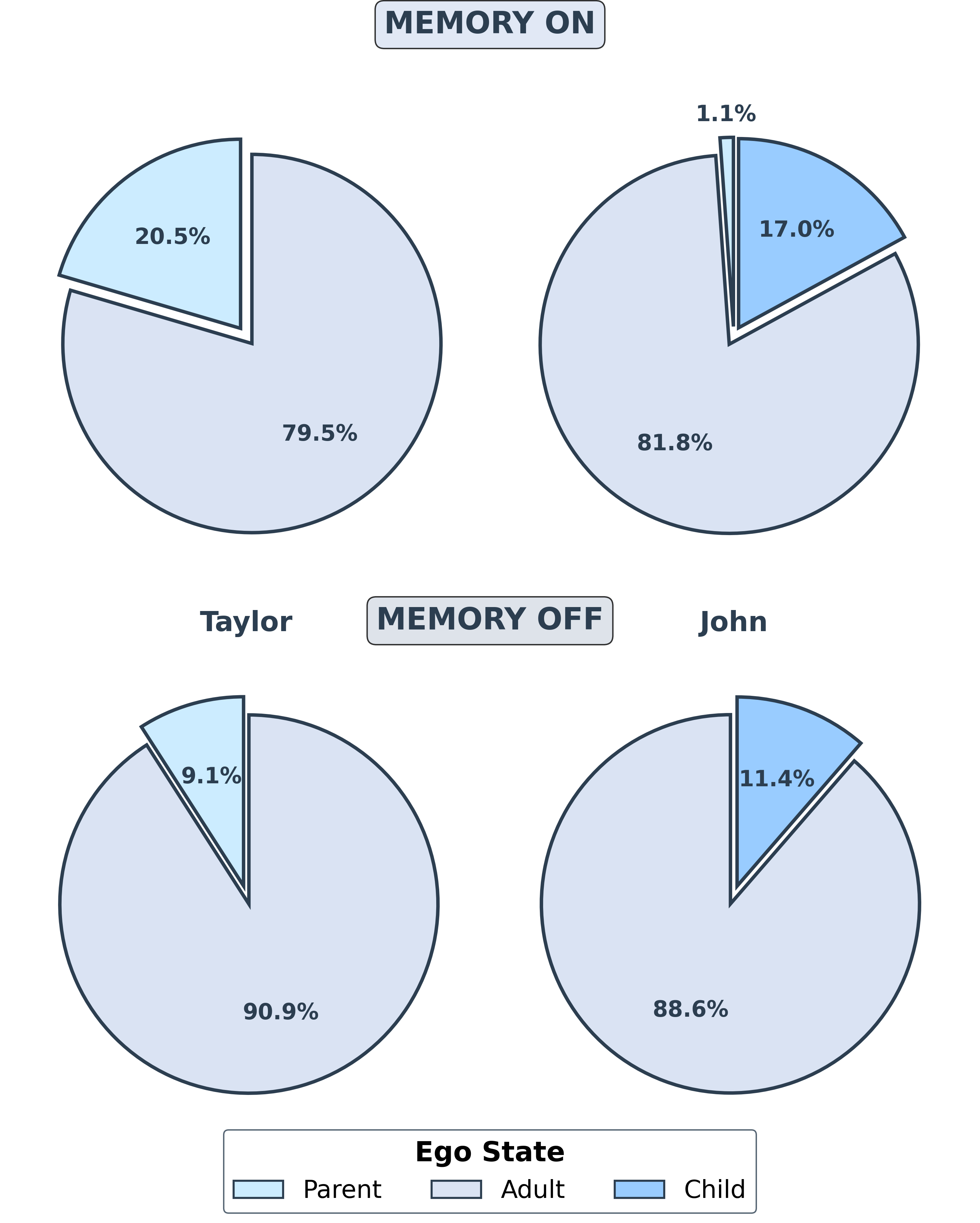}
  \caption{Distribution of Ego State Selection, Parent (P), Adult (A), and Child (C), for Agents Taylor (left column) and John (right column). The top row shows the results for the Memory ON condition, where agents had access to contextual information from their memory banks. The bottom row shows the results for the Memory OFF condition.}
  \label{fig5}
\end{figure}

The second condition, \textbf{Memory ON}, involves agents utilizing their information retrieval tools (see Section~\ref{sec:memory_retrieval}). In this setup, each of the three \emph{ego state} agents (Parent, Adult, Child) for both John and Taylor has access to its dedicated memory system with predefined memory items.

\begin{figure*}[t]
  \centering
  \includegraphics[width=\textwidth]{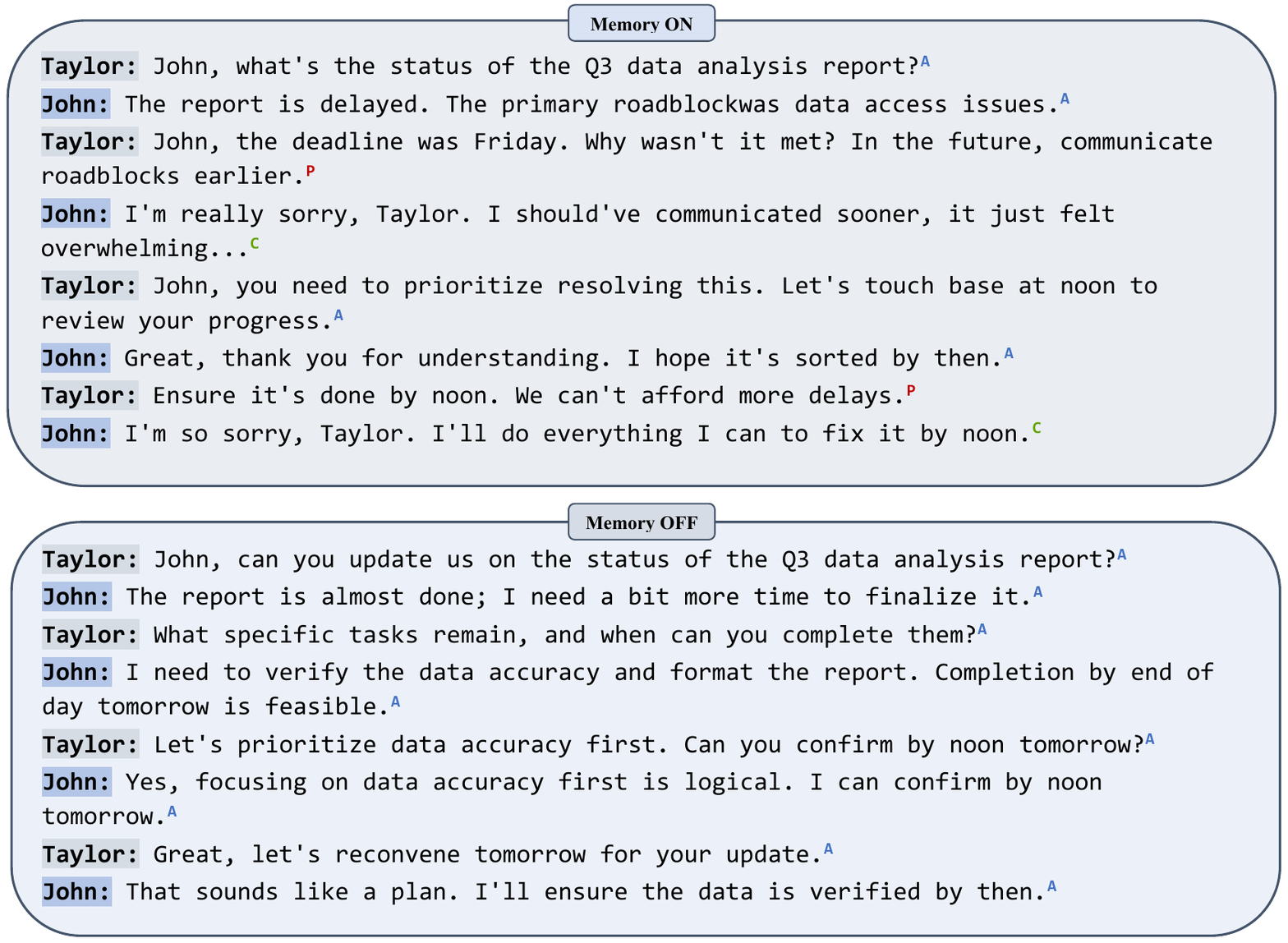}
  \caption{Comparative dialogues illustrating LLM agent behavior with information retrieval enabled (ON, top) versus disabled (OFF, bottom). \emph{Ego state} activations (P: Parent, A: Adult, C: Child) are indicated for each statement.}
  \label{fig4}
\end{figure*}

\subsection {Key Findings}
A primary observation is that the explicit modeling of Parent, Adult, and Child \emph{ego states} allowed agents to move beyond a default, often rational or solution-oriented, LLM behavior. Even in the \textbf{Memory OFF} condition, agents demonstrated the capacity to generate responses reflecting distinct \emph{ego state} perspectives. However, when memory retrieval was enabled (\textbf{Memory ON}), agent responses became more nuanced, consistent with their character, and emotionally grounded (see comparative dialogue examples in Figure~\ref{fig4}). 

Across both \textbf{Memory ON} and \textbf{OFF} conditions, the Adult \emph{ego state} was frequently selected by the meta-decision process for both Taylor and John. This is likely influenced by the professional work setting of the simulated scenario, where rational communication (characteristic of the Adult state) is often the expected norm. This indicates that while the architecture supports diverse \emph{ego state} expression, the conversational context and the nature of the task heavily influence which \emph{ego state} is chosen.

The ablation study (\textbf{Memory ON} vs. \textbf{OFF}) highlighted that access to contextual information influenced the distribution of selected \emph{ego states}. Specifically, in the \textbf{Memory ON} condition, John exhibited an increase in Child \emph{ego state} responses (from 10 to 15 of his turns), while Taylor's engagement from her Parent more than doubled (from 8 to 18) (see Figure~\ref{fig5} for detailed distributions). This shift suggests that the retrieved information provides stronger, more specific cues for producing a response that is more psychologically consistent and grounded. Such a memory-enhanced response becomes a much more compelling candidate for the meta-decision LLM, as it better aligns with the agent's core \emph{life script}. This leads to a higher selection rate of non-Adult \emph{ego states} and more dynamic interactions.

The increased activation of non-Adult states directly fostered the conditions for a more frequent Parent-to-Child dynamic. For instance, when Taylor communicated from her Parent, her messages were inherently more critical and evaluative. This type of input is a trigger for John's Child state, whose \emph{life script} is centered on feelings of inadequacy. This pattern, where a change in an \emph{ego state} by one agent prompts a complementary \emph{ego state} shift in the other, was also observable in \textbf{Memory OFF} condition, but less frequently due to the limited diversity of ego state selections. Future research will aim to make such \emph{complementary} (and \emph{crossed}, see Section~\ref{sec:ta_for_structuring_behavior}) \emph{transactions} more explicit within the simulation's logic and analysis. 

\section{Discussion}
While the initial results from applying the architecture are promising, we acknowledge that this research contains certain limitations which provide directions for future research. The current evaluation focuses mainly on qualitative analysis within a single dialogue scenario. This approach restricts how broadly these conclusions can be applied in different types of social interaction. Another important limitation concerns the memory component - these were predefined rather than developed through interaction experiences.

Based on observations, next research efforts will target multiple important areas to improve the proposed system and overcome current shortcomings. We intend to include more Transactional Analysis (TA) concepts like \emph{discounting} (see Section \ref{sec:ta_for_structuring_behavior}), \emph{strokes} (small units of recognition that satisfy the need to be noticed), and \emph{stamp collecting} patterns (accumulation of negative emotions) to make agent interactions more psychologically realistic \cite{Stewart12}. The next essential step involves the development of a more transparent, algorithmic mechanism to replace the current LLM-based selection of the final response. This new mechanism could incorporate a weighting system, where the answer is influenced by real-time conversational metrics, like accumulated 'emotional stamps' leading to a build-up of frustration. The final response could then be generated as a fusion of \emph{ego state} outputs, with each contribution proportional to its calculated weight, which would better simulate the internal psychological conflicts of human decision-making.

For improving memory functionality, we want to investigate approaches that enable agents to automatically generate and modify their \emph{ego state} memories during conversations. This might involve using reinforcement learning techniques to determine what experiences should be remembered and how these memories affect future responses. Most importantly, conducting broader testing with different scenarios and possibly including TA practitioner judges will be necessary to properly evaluate the advantages and complexities of this psychology-based agent design.

\section{Conclusion}
This paper has presented a novel approach to enhance the psychological realism of LLM-based agents through the integration of Transactional Analysis theory with contextual information retrieval mechanisms. The experimental evaluation demonstrates that modeling agents as composite systems of Parent, Adult, and Child \emph{ego states} leads to more nuanced and psychologically grounded interactions compared to traditional LLM agents. The ablation study reveals that memory-enabled agents exhibit more diverse \emph{ego state} activations. While the initial results are promising, several limitations are acknowledged including single scenario validation and reliance on predefined memory content, which present opportunities for future research. The implications of this research extend beyond technical improvements to LLM agents. Grounding agent behavior in established psychological theory opens new possibilities for applications in social science research, educational simulations, and therapeutic contexts.

\section*{Acknowledgments}
The work reported in this paper was partly supported by the Polish National Science Centre under grant 2024/06/Y/HS1/00197.

\bibliography{main}

\end{document}